\documentclass[aps,prb,twocolumn,showpacs]{revtex4}
\usepackage{pstricks}
\usepackage{graphicx}
\usepackage{epsfig}

\begin{document}

\title{Lowering Effective Coordination Promotes Adsorption of NO on Rh(100) and Rh/MgO(100)surfaces}

\author{Raghani Pushpa$^1$} 
\author{Prasenjit Ghosh$^{2,3}$} 
\author{Shobhana Narasimhan$^2$}
\author{Stefano de Gironcoli$^1$}
%\email{ shobhana@jncasr.ac.in}
\affiliation{$^1$Scuola Internazionale Superiore di Study Avanzati (SISSA),
Via Beirut 2/4, I-34014 Trieste, Italy}
\affiliation{$^2$Jawaharlal Nehru Centre for Advanced Scientific Research,  
Jakkur PO, Bangalore 560 064, India}
\affiliation{$^3$The Abdus Salam International Centre for Theoretical Physics (ICTP), Strada Costiera 11, I-34014 Trieste, Italy}

\begin{abstract}

We have studied the adsorption of NO, and the coadsorption of N and O, on four physical 
and hypothetical systems: unstrained and strained Rh(100) surfaces 
and monolayers of Rh atoms on strained and unstrained MgO(100) surfaces. We find that as we
go from Rh(100) to Rh/Mg0(100), via the other two hypothetical systems, the effective
coordination progressively decreases, the $d$-band narrows and its center shifts closer to
the Fermi level, and the strength of adsorption and co-adsorption increases.
Both strain and the presence of the oxide substrate contribute significantly to this. However,
charge-transfer is found to play a negligible role, due to a cancelling out between donation
and back-donation processes. Our results suggest that lowering effective coordination of Rh catalysts by strain,
roughening or the use of inert substrates might improve reaction rates for the reduction of
NO to N$_2$.

\end{abstract}


\maketitle

%%%%%%%%%%%%%%%%%%%%%%%%%%%%%%%%%%%%%%%%%%%%%%%%%%%%%%%%%%%%%%%%%%%%%%
\section{Introduction}

Though catalysts are crucial to the operation of many industrial and commercial processes, our understanding of the factors that make a good catalyst is still incomplete. Both the electronic and geometric structures of the catalyst are known to be important, and in recent years, it has become evident that coordination number can have a large effect on catalytic activity. The general understanding is that lower coordination leads to higher activity, though there are exceptions.\cite{lof213}  Industrial catalysts usually consist of small particles supported on a substrate. The presence of the substrate introduces additional factors, such as charge transfer between the catalyst and the support, a change in the local environment of catalyst atoms, and geometrical strain imposed by the presence of the substrate.

In this paper, we have studied some of these issues by performing calculations to study the adsorption of NO, as well as the coadsorption of N and O, on a variety of realistic and hypothetical Rh(100) surfaces. These configurations constitute the initial and final states in the dissociation of NO, which is a  crucial step in the reduction of nitrogen monoxide to nitrogen, for example in three-way catalysts in automobiles.  Rh is perhaps the best catalyst for this process,\cite{lof108} but it is also by far the most expensive precious metal. It is therefore desirable to gain a better understanding of what makes a good catalyst, so as to guide one in developing new catalysts that use less or no Rh. 

Towards this end, and in an attempt to gauge the contributions made by the chemical nature of the substrate, as well as strain, we have studied adsorption and coadsorption on four kinds of (100) surfaces: (i) the surface of an unstrained Rh crystal; (ii) the surface of a Rh crystal that has been stretched (expanded) so as to be commensurate with an MgO(100) surface; (iii) a monolayer of Rh on Mg0(100), with both the Rh and the substrate's in-plane spacing fixed to be that of Rh(100); and (iv) a monolayer of Rh on MgO(100), with all in-plane spacings fixed to be that of MgO(100). Note that straining systems (i) and (iii) leads to systems (ii) and (iv) respectively, while changing the substrate from Rh to MgO converts systems (i) and (ii) to systems (iii) and (iv) respectively. Qualitatively, it seems apparent that as one proceeds from system (i) to system (iv), the surface Rh atoms become progressively less coordinated; this idea is put on a quantitative footing further below!
 . 

Another factor -- magnetism -- seems likely to play a role in these systems. Bulk Rh is ``almost" ferromagnetic. There has been a long-running controversy over
whether or not the Rh(100) surface is magnetic;\cite{Morrison,SCWu,ChoSchefler,Nayak,Cho} 
however it seems clear that Rh monolayers
and clusters are magnetic.\cite{Eriksson,Zhu,Wu-Freeman,Reddy} This raises the question of what possible role magnetism may play in the operation of Rh catalysts; this issue is also dealt with in this paper. 

Though the adsorption of NO on Rh(100) has been studied by previous authors,
\cite{lof108,lof115,lof213} we
are not aware of any systematic programme of calculations that is similar in spirit to ours.

\section{Method}

Our calculations have been performed using {\it ab initio} density functional theory, using
the PWscf package of the Quantum-ESPRESSO distribution.\cite{QE} The spin-polarized version of the Kohn-Sham equations were
solved using ultrasoft pseudopotentials \cite{uspp} and a plane-wave basis with a cut-off 
of 30 Ry. Exchange and correlation effects
were treated using the Generalized  
Gradient Approximation (GGA) in the form suggested by Perdew, Burke and Ernzerhof.\cite{PBE}
In order to improve convergence, a Methfessel-Paxton smearing \cite{MPsmear} 
with a width of 0.03 Ry was used. 

Most results for the Rh(100) and Rh/MgO(100) slabs were obtained by using a
 $(1\times 1)$ asymmetric supercell containing four Rh layers, of which the outermost two (towards the adsorbate) were allowed to relax, while the inner two
were kept fixed at the appropriate bulk separation. However, some tests were also performed with symmetric slabs containing eight layers of Rh atoms.
Brillouin zone integrations for such $(1 \times 1) $ surface cells 
were carried out using a ($12\times 12 \times 1)$  
Monkhorst-Pack mesh.\cite{monkh}

The adsorption and co-adsorption studies were carried out using slabs with larger unit cells,
[viz., $(2\times 2)$ and $(2\times 3)$], together with corresponding k-point meshes  
[$(6\times 6 \times 1)$ and $(6 \times 4 \times 1)$ respectively].
For systems (i) and (ii), we found that a slab containing four Rh layers was sufficient
to give well-converged adsorption energies; however, for one particular adsorption geometry,
we found it necessary to use a slab with five Rh layers. All the results presented below
for adsorption and co-adsortpion on systems (iii) and (iv) were obtained using a slab 
containing one layer of Rh atoms over four MgO layers.

\begin{figure}
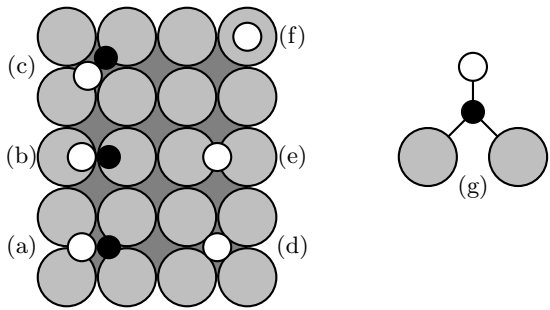

\caption{Schematic diagrams showing the geometries considered for
adsorption studies. The grey, white and black circles represent Rh, O and N atoms
respectively. The NO occupies different sites, and is oriented differently, in the
six cases depicted here. (a),(b) and (c) show top views of configurations where the NO sits
horizontally on the Rh substrate, while (d), (e) and (f) show top views of
configurations where the NO sits vertically on the substrate; in these, the N atoms
are not visible, as they sit directly below the O atoms. 
The picture (g) shows a side view of configuration (e).
}
\vspace{40mm}
\psset{unit=1mm}
\rput(-30,0){
\psframe[fillstyle=solid,fillcolor=gray](0,0)(24,32)
\multips(0,0)(0,8){5}{\multips(0,0)(8,0){4}
{\pscircle[fillstyle=solid,fillcolor=lightgray](0,0){4}}}
\pscircle[fillstyle=solid,fillcolor=white](2,4){2}
\pscircle[fillstyle=solid,fillcolor=black](5.6,4){1.6}
\pscircle[fillstyle=solid,fillcolor=white](2,16){2}
\pscircle[fillstyle=solid,fillcolor=black](5.6,16){1.6}
\pscircle[fillstyle=solid,fillcolor=white](2.8,26.8){2}
\pscircle[fillstyle=solid,fillcolor=black](5.2,29.2){1.6}
\pscircle[fillstyle=solid,fillcolor=black](20,4){1.6}
\pscircle[fillstyle=solid,fillcolor=white](20,4){2}
\pscircle[fillstyle=solid,fillcolor=black](20,16){1.6}
\pscircle[fillstyle=solid,fillcolor=white](20,16){2}
\pscircle[fillstyle=solid,fillcolor=black](24,32){1.6}
\pscircle[fillstyle=solid,fillcolor=white](24,32){2}
\psline(54,28)(54,22)(48,16)
\psline(54,22)(60,16)
\pscircle[fillstyle=solid,fillcolor=lightgray](48,16){4}
\pscircle[fillstyle=solid,fillcolor=lightgray](60,16){4}
\pscircle[fillstyle=solid,fillcolor=black](54,22){1.6}
\pscircle[fillstyle=solid,fillcolor=white](54,28){2}
\rput(-6,4){(a)}
\rput(-6,16){(b)}
\rput(-6,28){(c)}
\rput(30,4){(d)}
\rput(30,16){(e)}
\rput(30,32){(f)}
\rput(54,12){(g)}
%\rput(4,38){(A)}
%\rput(20,38){(B)}
}
\label{NO-Rh100}
\end{figure}

\begin{figure}
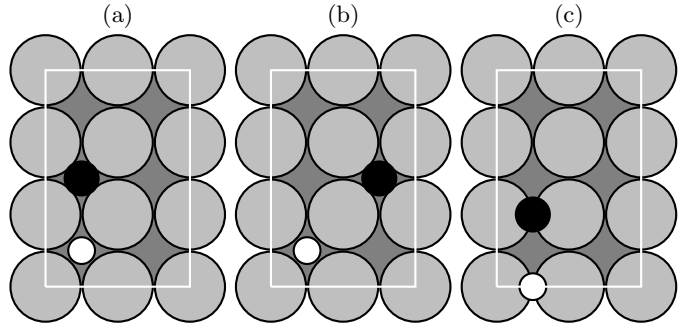

\caption{Top view of the geometries considered for N and O adsorption 
on Rh(100) surfaces. The grey, white and black circles represent Rh, O and N atoms
respectively.  
In configuration (a), N and O are at nearest neighbour hollow sites,
in (b) N and O  are at diagonally opposite hollow sites. In 
configuration (c), N and O are at the bridge sites between two Rh 
atoms. The white rectangle indicates the 2$\times$3 unit cell used for 
co-adsorption studies.}
\psset{unit=1.2mm}
\rput(-35,-30){
\psframe[fillstyle=solid,fillcolor=gray](0,0)(16,24)
\multips(0,0)(0,8){4}{\multips(0,0)(8,0){3}
{\pscircle[fillstyle=solid,fillcolor=lightgray](0,0){4}}}
\pscircle[fillstyle=solid,fillcolor=white](4,4){1.6}
%\pscircle[fillstyle=solid,fillcolor=black](12,12){2.0}
\pscircle[fillstyle=solid,fillcolor=black](4,12){2.0}
%\pscircle[fillstyle=solid,fillcolor=black](12,12){2.0}
%\psline[linecolor=white](0,0)(0,16)(16,16)(16,0)(0,0)
\psline[linecolor=white](0,0)(0,24)(16,24)(16,0)(0,0)
%\rput(20,12){(a)}
\rput(8,30){(a)}
%%\rput(20,12){(b)}
}
\rput(-10,-30){
\psframe[fillstyle=solid,fillcolor=gray](0,0)(16,24)
\multips(0,0)(0,8){4}{\multips(0,0)(8,0){3}
{\pscircle[fillstyle=solid,fillcolor=lightgray](0,0){4}}}
\pscircle[fillstyle=solid,fillcolor=white](4,4){1.6}
%\pscircle[fillstyle=solid,fillcolor=black](12,12){2.0}
%\pscircle[fillstyle=solid,fillcolor=black](4,12){2.0}
\pscircle[fillstyle=solid,fillcolor=black](12,12){2.0}
%\psline[linecolor=white](0,0)(0,16)(16,16)(16,0)(0,0)
\psline[linecolor=white](0,0)(0,24)(16,24)(16,0)(0,0)
%\rput(20,12){(a)}
%%%\rput(-4,12){(b)}
\rput(8,30){(b)}
%%\rput(20,12){(b)}
}
\rput(15,-30){
\psframe[fillstyle=solid,fillcolor=gray](0,0)(16,24)
\multips(0,0)(0,8){4}{\multips(0,0)(8,0){3}
{\pscircle[fillstyle=solid,fillcolor=lightgray](0,0){4}}}
\pscircle[fillstyle=solid,fillcolor=white](4,0){1.6}
\pscircle[fillstyle=solid,fillcolor=black](4,8){2.0}
%\pscircle[fillstyle=solid,fillcolor=black](0,12){2.0}
\psline[linecolor=white](0,0)(0,24)(16,24)(16,0)(0,0)
%%\rput(21,8){(c)}
\rput(8,30){(c)}
}
\vspace{35mm}
\label{N+O-Rh100}
\end{figure}

In adsorption studies, six different geometries were considered; these
are shown in Fig.~\ref{NO-Rh100}, while Fig.~\ref{N+O-Rh100} shows the three different geometries for the 
co-adsorption studies. These two figures show
(schematically) the geometry prior to relaxation; after relaxation (in
the absence of symmetry constraints) the geometries remained roughly similar, though bond lengths changed. 

%%%%%%%%%%%%%%%%%%%%%%%%%%%%%%%%%%%%%%%%%%%%%%%%%%%%%%%%%%%%%%%%%%%%%%%
%%%%%%%%%%%%%%%%%%%%%%%%%%%%%%%%%%%%%%%%%%%%%%%%%%%%%%%%%%%%%%%%%%%%%%%%%%%

%%%%%%%%%%%%%%%%%%%%%%%%%%%%%%%%%%%%%%%%%%%%%%%%%%%%%%%%%%%%%%%%%%%%%%%

%%%%%%%%%%%%%%%%%%%%%%%%%%%%%%%%%%%%%%%%%%%%%%%%%%%%%%%%%%%%%%%%%%%%%%%

As we will show in the next section, in cases (ii), (iii) and (iv), the Rh(100) surface was
found to be magnetic. In such cases, one needs to worry about the possibility that the bottom surface acquires a magnetic moment and is thus no longer representative of a bulk-like layer. To mitigate this, in
these cases we deposited a layer of H atoms on the bottom surface, which has the effect
of quenching magnetization on that surface.

\section{Results}

\subsection {Preliminary Tests: bulk Rh, bulk MgO and gas-phase NO}

We obtained a value of 3.85 \AA\ for the lattice constant
of bulk Rh, which is in excellent agreement with the experimental value of 3.80 \AA 
\cite{Ashcroft}
and previous theoretical values of 3.87 \AA.\cite{bondino} Our value implies that the 
distance between
nearest-neighbor (NN) Rh atoms on the (100) surface is  2.73 \AA, while the interlayer distance
between bulk-like layers in a Rh(100) slab is 1.93 \AA.

For MgO, we obtained a lattice constant of 4.25 \AA, which is identical to the value obtained 
in previous calculations,\cite{nokbin} and in good agreement with the experimental value
of 4.21 \AA.\cite{wykoff} Note that this would imply that if a Rh monolayer were to be deposited
commensurately with an MgO(100) substrate, the NN Rh-Rh distance within the monolayer would
be increased to 3.00 \AA, which corresponds to a strain of 9.9\%. 

For NO in the gas phase, we obtained a binding energy of 7.13 eV, and an N-O bond length
of 1.17 \AA\ . For comparison, the experimental values are 6.5 eV and 1.15 \AA\ respectively.
\cite{Johnson}

In all these cases, it can be seen that our results are in good agreement with both experiments
and previous calculations, lending support to the validity of our approach.

\subsection{Case (i) - Unstrained Rh(100) surface}

For the surface energy of the clean and unstrained Rh(100) surface [case (i)], we obtain a value of 
1.12 eV per surface atom, which agrees exactly with the value obtained in a previous
calculation,\cite{bondino} and is also in reasonably good agreement with the experimental
value of 1.27 eV per surface atom.\cite{teeter} We find that the first interlayer
spacing $d_{12}$ is contracted by 3.5\% with respect to the bulk interlayer spacing of
1.93 \AA;
this is similar to the contraction of 4.0\% found in an earlier study,\cite{bondino}
but more than the contraction of 1.4\% $\pm$ 1.4\% reported experimentally.\cite{teeter}
We find that the next two interlayer spacings, $d_{23}$, and $d_{34}$, are both expanded
by 0.77\%; the next interlayer spacing $d_{45}$ is very close to the bulk interlayer spacing.
 We find that the surface is
not ferromagnetic, which is in agreement with several previous studies,\cite{Nayak,Cho} though it
disagrees with some reports in the literature.\cite{Morrison,SCWu}

Following previous authors,\cite{Laura} we define an effective coordination number for
the surface as:

\begin{equation}
n_e = \Sigma_j\rho_{Rh}^{at}(R_{ij})/\rho_{Rh}^{at}(R_{bulk})
\end{equation}
where,
$\rho_{Rh}^{at}(R)$ is the atomic charge density of an isolated Rh atom 
as a function of $R$, the distance from the nucleus, and the sum is taken over all the nearest neighbour
atoms $j$ around a surface Rh atom $i$. $R_{ij}$ is the distance between atoms 
$i$ and $j$, and $R_{bulk}$ is the NN distance for bulk Rh. For case (i), we find
$n_e$ = 8.49. This is slightly increased with respect to the nominal surface coordination (the number
of nearest neighbor atoms for a surface atom) of 8, due primarily to the contraction of $d_{12}$ relative
to the bulk interlayer spacing.

\begin{figure}
\epsfig{file=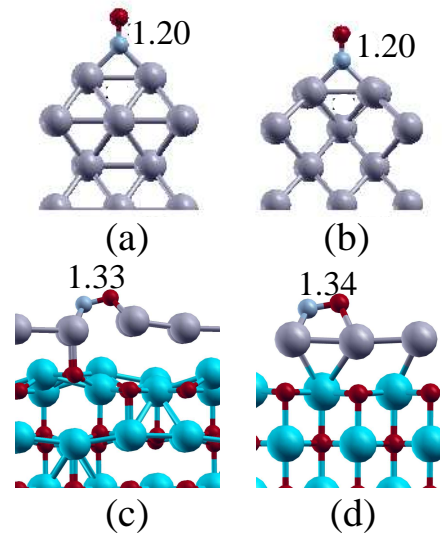,width=60mm,clip=}
\caption{(In color online). The most stable configurations for adsorbed NO in  (a) case (i), (b) case (ii),
(c) case (iii), and (d) case (iv). In cases (i) and (ii), vertical configuration
of NO at a bridge site is the most stable; whereas in cases (iii) and (iv),
horizontal configuration at the hollow site is the most stable. Small blue and red, and large grey and turquoise
spheres represent N, O, Rh and Mg atoms respectively.}
\label{NO-4cases}
\end{figure}

We define the adsorption
energy as $E_{ads} = E_{NO:Rh(100)} - E_{Rh(100)} - 
E_{NO}^{gas}$, where $E$ is the value obtained from {\it ab initio} calculations for
the total energy of the corresponding configuration. In agreement with previous results,
\cite{lof108,bondino} we find that the most favorable adsorption geometry on an unstrained Rh(100) slab
is the ``vertical bridge" (VB) [corresponding to Figs.\ref{NO-Rh100}(e) and (g); the adsorption
geometry is also shown in Fig.~\ref{NO-4cases}(a)]. With a $(2 \times 2)$ unit
cell (i.e., a coverage of 1/4 ML), we obtain
$E_{ads}$ = -2.59 eV in this geometry; and the N-O bond length is increased slightly from
the gas-phase value of 1.17 \AA \ to 1.20 \AA. The next most favorable adsorption geometry is the ``horizontal hollow" (HH) [see Fig.\ref{NO-Rh100}(a)]
with $E_{ads}$ = -2.47 eV, and an N-O bond length of 1.31 \AA \ in
the latter geometry. The details of the adsorption geometries
for both VB and HH are given in Table \ref{adsgeom}.
All these values are in excellent agreement with a 
previous study.\cite{bondino} 

%%%%%%%%%%%%%%%%%%%%%%%%%%%%%%%%%%%%%%%%%%%%%%%%%%%%%%%%%%%%%%%%%%%%%%%
%%%%%%%%%%%%%%%%%%%%%%%%%%%%%%%%%%%%%%%%%%%%%%%%%%%%%%%%%%%%%%%%%%%%%%

We have also investigated the geometry and energetics of the co-adsorption of N and O on
this surface. We find that both N and O prefer to occupy four-fold hollow sites, as has also
been reported in a previous paper.\cite{lof115} In Table \ref{unstrnd-N+O}, we show
how the co-adsorption energy, defined as $E_{coads} = E_{N+O:Rh(100)} - E_{Rh(100)} - 
E_{NO}^{gas}$,
varies with adsorbate coverage and adsorption geometry. We find that the N and O atoms prefer
to occupy next-nearest-neighbor hollow sites over nearest-neighbor hollow sites. Also,
the magnitude of $E_{coads}$ increases as the coverage is decreased. These  observations suggest
that there is a repulsive interaction between the N and O atoms after dissociation of NO. 
We note that in a previous study,\cite{lof115}
a coadsorption energy of -3.50 eV was obtained at
a coverage of 1/16 ML, which is in keeping with our conclusions here.

%%%%%%%%%%%%%%%%%%%%%%%%%%%%%%%%%%%%%%%%%%%%%%%%%%%%%%%%%%%%%%%%%%%%%%
\begin{table*}
\caption{The Rh-Rh bond length ($d_{Rh-Rh}$), the Rh-N bond length ($d_{Rh-N}$), the Rh-O bond length ($d_{Rh-O}$)
and the N-O bond length ($d_{N-O}$) of NO on different adsorption
sites for the four cases. The number given below each case denotes the Rh-Rh
in plane distance for the clean surfaces. $Rh_N$ and $Rh_O$ indicate the distances between two
Rh atoms bonded to N and O atoms respectively. All distances are given in \AA.}
\begin{center}
\begin{tabular*}{5.5in}{c@{\extracolsep{\fill}}c@{\extracolsep{\fill}}
c@{\extracolsep{\fill}}c@{\extracolsep{\fill}}c@{\extracolsep{\fill}}
c@{\extracolsep{\fill}}}
\hline
System & Site and geometry & $d_{Rh-Rh}$ & $d_{Rh-N}$ & $d_{Rh-O}$& $d_{N-O}$\\
\hline
case (i) & VB & 2.74 & 1.96 & - & 1.20 \\
(2.73) & HH & 2.75 (Rh$_N$)& 1.98 & 2.21 & 1.31 \\
       & & 2.79 (Rh$_O$) & & & \\
\hline
case(ii)& VB & 2.70 & 1.96 & - & 1.20\\
(3.00) & HH & 3.13 (Rh$_N$) & 1.98 & 2.22 & 1.32 \\
& & 3.21 (Rh$_O$) & & & \\
\hline
case (iii)& VB & 2.75 & 1.93 & - & 1.20 \\
(2.73)& HH & 2.70 (Rh$_N$) & 1.95 & 2.15 & 1.33 \\
& & 2.95 (Rh$_O$)& & & \\
\hline
case(iv)& VB & 2.56 & 1.94 & - & 1.21\\
(3.00)& HH & 2.66 (Rh$_N$) & 1.95 & 2.18 & 1.34\\
&  & 2.59 (Rh$_O$) & & & \\
\hline
\end{tabular*}
\end{center}
\label{adsgeom}
\end{table*}

%%%%%%%%%%%%%%%%%%%%%%%%%%%%%%%%%%%%%%%%%%%%%%%%%%%%%%%%%%%%%%%%%%%%%
%%%%%%%%%%%%%%%%%%%%%%%%%%%%%%%%%%%%%%%%%%%%%%%%%%%%%%%%%%%%%%%%%%%%%%%
\begin{table}
\caption{Dependence of coadsorption energy $E_{coads}$ on coverage and adsorption geometry.
The first row corresponds
to a coadsorption of N and O in a 2$\times$2 unit cell in next-nearest-neighbor
hollow sites, while the second and third rows (a) correspond to
results for the geometries depicted in Figs.\ref{N+O-Rh100}(a) and (b)
respectively. Here,
C = coverage of N = coverage of O.}

\vspace{1cm}
\noindent\begin{tabular}{|c|c|c|c|}                  \hline
 Coverage C   & $E_{coads}$ (eV)  &  $r_{N-Rh}$(\AA) & $r_{O-Rh}$ (\AA) \\ \hline
 1/4 ML       &  -3.20              & 2.03           & 2.14            \\
 1/6 ML (a)   &  -3.10              & 1.99, 2.10     & 2.04, 2.35      \\
 1/6 ML (b)   &  -3.34              & 2.03, 2.04     & 2.14, 2.17      \\ \hline
\end{tabular}
\vspace{1cm}
\label{unstrnd-N+O}
\end{table}
%%%%%%%%%%%%%%%%%%%%%%%%%%%%%%%%%%%%%%%%%%%%%%%%%%%%%%%%%%%%%%%%%%%%
\subsection{Case (ii) - Stretched Rh(100) surface}

When an in-plane expansion of 9.9\% is imposed on a (1$\times$1) 
Rh(100) slab, we find that the bulk interlayer spacing ($\approx d_{45}$) is reduced
by 11.3\% with respect to the unstrained case and becomes
equal to 1.71 \AA.  Upon allowing all interlayer distances to relax in an
eight-layer slab, the first three interlayer spacings $d_{12}$, $d_{23}$ and
$d_{34}$ are reduced to 1.59, 1.76 and 1.73 \AA \ respectively.  These are
contracted significantly, with respect to the bulk interlayer distance in
unstrained Rh(100), by 17.3, 8.9 and 10.5 \% respectively. The net effect of
having longer intralayer distances but shorter interlayer distances at the
surface is that now the effective coordination $n_e$ = 6.65, i.e., as a result of
stretching, the effective coordination has decreased on going from case (i) to
case (ii).

\begin{figure} \epsfig{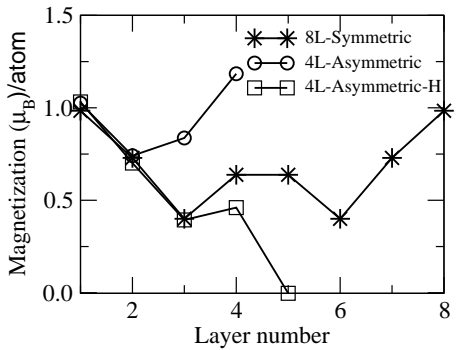}
\caption{Magnetization of the stretched Rh(100) surface [case (ii)] for an
eight-layer symmetric slab (stars), a four-layer asymmetric slab (cirlces) and
a four-layer asymmetric slab with H adsorbed at the bottom of the slab
(squares).} \label{Magvsatm-4L-8L-H} \end{figure}

In Fig.~\ref{Magvsatm-4L-8L-H}, we show the layer-resolved magnetization per
atom of a symmetric (1$\times$1) stretched eight-layer slab (stars).
Interestingly, the stretched Rh(100) surface is now magnetic with surface
magnetization of 1.0 $\mu_B$/atom.  In order to reduce computational load, it
is a common practice to consider asymmetric slabs where some of the bottom
layers are fixed at their bulk positions. The circles in Fig.
\ref{Magvsatm-4L-8L-H} show the magnetization per atom in a (1$\times$1)
four-layer slab with the bottom two layers fixed at 1.71 \AA\ interlayer
spacing and the top two layers relaxed.  We find that only in the top two
layers is the magnetization close to the one in the  layers of the symmetric
slab, and a ``spurious" magnetization as large as 1.2 $\mu_B$/atom is present
on the frozen-geometry bottom surface. In order to quench this spurious
magnetization, we considered also an asymmetric four-layer Rh slab where H
atoms were adsorbed on the bottom surface.  The results are again reported in
Fig. \ref{Magvsatm-4L-8L-H} as squares, and we now find that the magnetic
moments on all layers are very similar for the four-layer and eight-layer
slabs, giving us confidence that adsorption energies obtained with the former
will be accurate. 

The interlayer distances are also very similar in the eight-layer symmetric
slab and in the four-layer slab with H adsorbed on the bottom surface, but are
(slightly but perceptibly) different on the four-layer slab without H on the
bottom surface.

Adsorbing H on the bottom surface also turns out to be crucial in correctly
predicting the stable adsorption geometry for NO. In the absence of H, it is
found that the VB and HH configurations are degenerate, with both having an
adsorption energy $E_{ads}$ = -2.83 eV; however, when H is adsorbed on the
bottom surface, the VB (with $E_{ads}$ = -2.86 eV) is found to be clearly
favored over the HH (with $E_{ads}$ = -2.71 eV); for the reasons stated above,
we believe that the set of numbers obtained with H atoms on the bottom surface
is more to be trusted.  This stable adsorption geometry is depicted in
Fig.~\ref{NO-4cases}(b).

In order to gauge the effects of magnetization on adsorption energies, we also
performed non-spin-polarized (NSP) calculations. We found that the magnitude of
$E_{ads}$ increases in the absence of magnetization; however, the difference
between the VB and HH geometries is maintained: with NSP calculations, the
former gives $E_{ads}=-2.96$ eV, while the latter gives $E_{ads} = -2.82$ eV. 
%
%%%%%%%%%%%%%%%%%%%%%%%%%%%%%%%%%%%%%%%%%%%%%%%%%%%%%%%%%%%%%%%%%%%%%%
%%%%%%%%%%%%%%%%%%%%%%%%%%%%%%%%%%%%%%%%%%%%%%%%%%%%%%%%%%%%%%%%%%%%%%

We found that NO adsorption leads to a significant distortion in the position
of Rh atoms (see Table \ref{adsgeom}). When NO is adsorbed in the VB configuration, the Rh-Rh distance
for the two Rh atoms bonded to N is reduced significantly, from 3.00 \AA \ to
2.70 \AA.  However, for the HH configuration, the
distances between the two Rh atoms bound to N and O increase to 3.13 and 3.21 \AA \ respectively.  The internal bond length in NO is increased in all cases; to 1.20
\AA \ for the VB and to 1.32 \AA \ for the HH configuration. Note that this
also suggests that it might be easier to break the N-O bond in the HH
configuration.

Next, we co-adsorb N and O on this surface. The co-adsorption geometries
considered by us are shown in Fig. \ref{N+O-Rh100}.  When we start from a
configuration where N and O are at NN hollow sites (see
Fig.\ref{N+O-Rh100}(a)), the O atom moves away from N, and sits at the bridge
site which is equidistant from the N atom and its image. This configuration represents the most favorable co-adsorption
geometry, with $E_{coads} = -3.47$ eV. The second most favourable configuration
is one where both N and O sit at NNN hollow sites (Fig.~\ref{N+O-Rh100}(b))); here  $E_{coads} = -3.33$ eV. Unlike in case (i),
the stretched Rh(100) surface also has a local minimum when both the
atoms sit at bridge sites (see Fig.~\ref{N+O-Rh100}(c)). This is the
configuration that the system assumes just after the dissociation of NO, and
corresponds to $E_{ads} = -3.24$ eV. As in case (i), the magnitude of $E_{ads}$
is increased by $\sim 0.3$ eV in the first two adsorption geometries upon
performing an NSP calculation; however, in the third case, $E_{ads}$ is
essentially unchanged.  It is also worth noting that changes in interatomic
distances are slight when comparing spin-polarized (SP) and NSP calculations.

\subsection{ Case (iii) - Monolayer of Rh on MgO(100), at Rh(100) lattice
constant}

The third case we consider is again an artificial one. We consider first an
MgO(100) substrate that has been strained (compressed) in-plane, so that it has
the same in-plane lattice constant as in case (i). As a result of this in-plane
compression, we find that the first two interlayer spacings are both increased
by 8.15\% 
relative to the interlayer spacing in unstrained MgO. There is also a
noticeable rumpling of Mg-O layers near the surface. We define the rumpling as
$\Delta z = (z_{O}-z_{Mg})/d_0$, where $z_O$ and $z_{Mg}$ are the
$z$-coordinates (normal to the surface plane) of O and Mg respectively, and
$d_0$ is the bulk interlayer spacing.  Then, the outermost MgO layer has
$\Delta z =2 .0\%$, while the second MgO layer has  $\Delta z =-0.1\%$.  Note
that the rumpling is opposite in the two outermost layers: in the topmost
layer, oxygen atoms are displaced further away from the substrate than Mg
atoms, whereas for the second layer, the reverse is true.

When a monolayer of Rh is deposited pseudomorphically on this compressed MgO
substrate, it binds with a binding energy $E_{bin}$ of -3.59 eV per Rh atom. Here, the binding energy has been defined by
$E_{bin}=E_{Rh:MgO}-E_{MgO}-E_{Rh}^{gas}$, where $E_{Rh:MgO}$ is the total
energy of the Rh/MgO slab, $E_{MgO}$ is the energy of the MgO slab alone, and
$E_{Rh}^{gas}$ is the energy of an isolated Rh atom in the gas phase. The Rh
atoms sit atop the oxygen atoms of the outermost MgO layer, with a Rh-O
separation of 2.22 \AA.  The deposition of the monolayer changes both the
rumpling and the relaxation of the outermost MgO layers.  The first two layers
now display a rumpling of -4.2 \% and 0.7\% respectively; note that the
deposition of Rh has actually reversed the direction of rumpling.  There is
also a change in interlayer spacings, with the distance between the two
outermost MgO layers now expanded by 9.55\% relative to the interlayer distance
in unstrained MgO, while the expansion of the next interlayer spacing is now reduced slightly, from the value of 8.15\% (in the absence of the Rh adlayer) to 8.05\%.
The Rh monolayer in this case is found to be magnetic,
with a magnetic moment of 1.5 $\mu_B$/atom.

\begin{figure} \epsfig{file=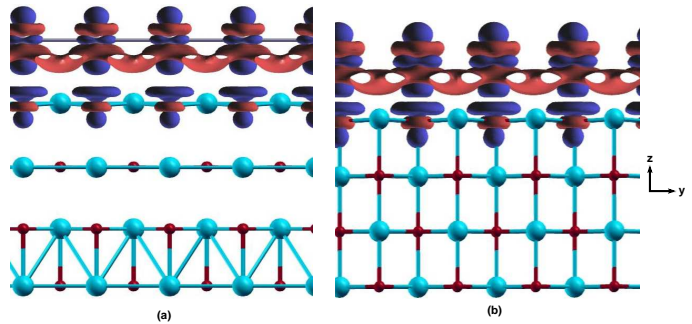 ,width=90mm,clip=} \caption{Charge
transfer on placing a monolayer of Rh atoms on an MgO substrate for (a) case
(iii) and (b) case (iv). Red and dark blue lobes represent regions where the
charge has increased and decreased respectively. Red and turquoise spheres
represent O and Mg atoms respectively, the grey Rh atoms are not visible
because they are surrounded by charge-transfer lobes. Note that both the Rh
overlayer and the O atoms in the topmost Mg-O layer display both red and blue
lobes.} \label{ctrans} \end{figure}

In principle, placing metal atoms on an oxide substrate can lead to charge
transfer. In Fig.~\ref{ctrans}(a), we have plotted the redistribution of charge
upon placing a layer of Rh atoms on an MgO substrate. In this figure, red and
dark blue regions represent areas where the charge has increased and decreased
respectively. It is interesting to note that one observes {\it both} red and
dark blue lobes on {\it both} the Rh layer and the topmost Mg-O layer; this is
because of the simultaneous presence of donation and back-donation between the
adsorbate (Rh) and the substrate (MgO), as is well-known for such systems.
Interestingly, these two processes essentially cancel out for this system, the
net charge transfer, as calculated by Lowdin population analysis, is only 0.015 electrons per Rh atom between the Rh
overlayer and the MgO substrate.

%%%%%%%%%%%%%%%%%%%%%%%%%%%%%%%%%%%%%%%%%%%%%%%%%%%%%%%%%%%%%%%%%%%%%%%
%%%%%%%%%%%%%%%%%%%%%%%%%%%%%%%%%%%%%%%%%%%%%%%%%%%%%%%%%%%%%%%%%%%%%%
The effective coordination $n_e$ is now lowered to 5.51. Note that $n_e$ is
lower for both cases (ii) and (iii) relative to case (i); however, the lowering
in case (iii) (which is due to the replacement of the Rh substrate by an MgO
support), is more significant than in case (ii) (where it is due to stretching
the system to the lattice constant of MgO). Similarly, the magnetic moment is
larger in case (iii) than in case (ii).

Upon adsorbing NO on this compressed Rh/MgO(100) system, we find that the most
favorable adsorption geometry is the HH [depicted in Fig.~\ref{NO-4cases}(c)],
with $E_{ads} = -3.38$ eV, which is larger in magnitude than the adsorption
energy of -3.17 eV for the VB configuration.
Note that this is in contrast to
the most favored adsorption geometry for both cases (i) and (ii); it is now
favorable for both N and O to bind to Rh atoms.

For the VB case, the distance
between the two Rh atoms bound to the N (Rh$_N$) increases slightly from 2.73 \AA\
to 2.75 \AA. On the contrary, for the HH geometry, the distance between
the same pair of Rh atoms decreases to 2.70 \AA, while that between the Rh
atoms attached to the O atom (Rh$_O$) increases to 2.95 \AA; this is because,
as a result of strong binding between NO and the Rh atoms, the Rh$_N$ pair is pulled
out of the surface. 

Moreover, upon adsorbing NO, the magnetization of the Rh atoms is quenched.
For the HH NO adsorption geometry, this reduction is as follows:
for the two Rh atoms to which N is attached, the reduction is by 74\%, for
the two Rh atoms to which O is attached, it is by 45\%, and for the two Rh
atoms which are furthest from NO, the reduction is only by 9\%.

Once again, we find
significantly enhanced binding upon performing an NSP calculation, with
$E_{ads} = -4.07 $ eV and -3.82 eV for the HH and VB geometries respectively.

The co-adsorption energies are -4.64 -3.32 eV when the N and O
atoms are in the NN hollow (Fig.~\ref{N+O-Rh100}(a)) and bridge
(Fig.~\ref{N+O-Rh100}(c)) sites respectively.

\subsection{Case (iv) - Monolayer of Rh on MgO(100), at MgO lattice constant}

Upon putting a pseudomorphic layer of Rh atoms on an unstrained MgO(100)
substrate, we find that Rh atoms again preferentially occupy the sites atop O
atoms. The binding energy for this configuration, defined as in case (iii), is
found to be $E_{bin}=-3.27$ eV per Rh atom; this is somewhat less in magnitude than the
value of -3.99 eV obtained in a previous
study.\cite{nokbin} The Rh-O distance is found to be 2.25 \AA, which is larger
than the value of 2.10 \AA \ reported earlier.\cite{nokbin} Both the Mg and O
atoms in the topmost layer move outwards, away from the substrate, so that the
mean interlayer distance between the top two MgO layers is increased by 0.73\%
relative to the bulk interlayer spacing. However, there is a considerable
rumpling (Mg atoms are higher up on the surface than oxygen atoms), with
$\Delta z =-4.0\%$, of the topmost MgO layer; this is in good agreement with one
earlier reported value of -4.3\%,\cite{Wu-Freeman} but somewhat less than the value of
-6.1\% obtained in another study.\cite{nokbin} The rumpling is reduced to +1.0\%
in the MgO layer below this one.  Note that for both layers the rumpling is reversed with respect to that
observed for a bare MgO(100) surface.

For a Rh atom in case (iv), we obtain $n_e$ = 3.59. Note that this is
significantly lowered with respect to the value of 8.49 obtained in case (i),
due to a combination of two factors: the replacement of substrate Rh atoms by
Mg and O atoms, and the stretching to the MgO lattice constant.

Once again, we find that the surface Rh atoms are magnetic, with a moment of
1.5 $\mu_B$/atom.

In Fig.~\ref{ctrans}(b), we present a plot of charge transfer for the Rh/MgO
system.  From this figure, it is evident that, as in case (iii), there is both
donation and back-donation between the overlayer Rh atoms and the O atoms in
the MgO substrate. Once again, these two processes effectively cancel out, and
there is a net charge transfer of only 0.018 electron per Rh atom from the oxygen to
the Rh atom.

Next, we study the adsorption of NO on this Rh/MgO(100) system.  As in case
(iii) above, we find that it is now most favorable for the NO atom to lie
horizontally on the surface: $E_{ads}=-4.08$ eV and -3.87 eV for the HH and VB
geometries respectively. In Fig.~\ref{NO-4cases} (d) we have depicted the
lowest-energy adsorption geometry, corresponding to the HH configuration.

With the adsorption of NO on 1 ML of Rh on MgO, the Rh atoms
to which the NO is attached come closer together for both the HH and VB
geometries (see Table \ref{adsgeom}). We speculate that this may be because
Rh-Rh bonds are under considerable tensile stress when deposited pseudomorphically
on MgO; the adsorption of NO breaks the symmetry and allows the Rh atoms bonded to
NO to come closer together.

As in case (iii), upon adsorbing NO, the magnetization of the Rh atoms is quenched.
Out of the six Rh atoms present in the unit cell, the spin polarization
of each of the two Rh atoms to which N is attached is negligible,
while those to which O is attached is reduced by about 81\% and the
spin polarization of the two Rh atoms which are furthest from the NO
is reduced by about 28\%. Similar effects have also been observed
previously, by us for NO adsorbed on small Rh clusters,\cite{NOclust} and 
by Hass \textit{et al.} in their studies
of NO adsorption on a hypothetical monolayer of Rh atoms.\cite{Hass} Apart from
the fact that the latter group of authors worked with a monolayer of Rh
(i.e., there was no substrate), there are other differences between our
calculations and theirs: they did the calculations for NO adsorbing in the
VB geometry alone, fixed the Rh-Rh distance at the value for bulk Rh,
and did not relax the coordinates of NO.
While they
found that the magnetic moments on all the atoms of the monolayer
are negligible, we find that the magnetism of those Rh atoms which
are attached to the NO molecule is very strongly suppressed, while
in the other Rh atoms there is a slight reduction.

As in the previous cases, the magnitude of $E_{ads}$ increases upon performing
an NSP calculation; one obtains values of -4.38 and -3.93 eV for the HH and VB
configurations respectively.

Upon co-adsorbing NO on this Rh/MgO slab, we find that it is preferable for
both the N and O to be at NN hollow sites rather than at NN bridge sites; the former
leads to a co-adsorption energy of -4.74 eV, while the latter leads to a
co-adsorption energy of -4.10 eV.  

\section {Discussion of Trends} \label{discussion} 
\begin{figure} \epsfig{file=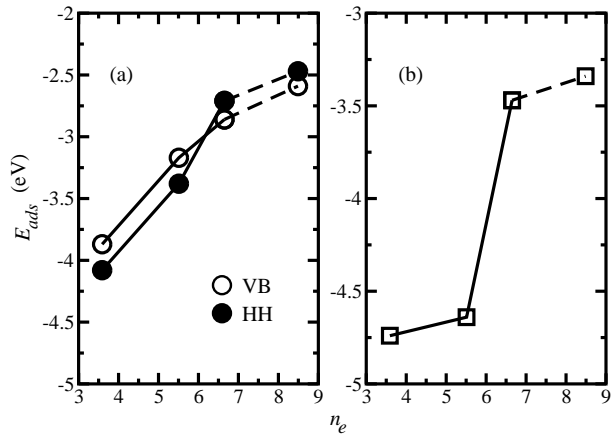 ,width=80mm,clip=}
\caption{Variation of (a) adsorption energy of NO, $E_{ads}$, and (b)
co-adsorption energy of N and O, with effective coordination $n_e$. The open
and filled circles in (a) represent the vertical bridge and horizontal hollow
adsorption geometries respectively.  The squares in (b) represent the
co-adsorption energy for the lowest energy configuration of all geometries
considered by us.} \label{eads_ecoads_vs_neff} \end{figure}

We have seen that as we go
from case (i) to case (iv), the effective coordination $n_e$ decreases
progressively and significantly. This decrease in effective coordination is
accompanied by (and, in our interpretation, causes) a significant increase in
the strength of adsorption of NO and co-adsorption of N and O. Our main results
are encapsulated in Fig.~\ref{eads_ecoads_vs_neff}. In the panel on the
left-hand-side, we have presented our results for the adsorption energy of NO
on these systems. The open circles represent results for the vertical bridge
geometry, and the filled circles those for the horizontal hollow geometry. The
following features are evident from this graph: (a) for a given geometry, the
magnitude of $E_{ads}$ increases monotonically as $n_e$ is decreased; (b) the
slope of this graph is larger for the HH than for the VB; this can be
rationalized as being due to the fact that for the HH, both the N and O are
bonded to Rh atoms, whereas for the VB, only N is bonded to Rh atoms; (c) as a
consequence of this the HH geometry becomes more favorable at lower $n_e$; (d)
the slope of the lines connecting cases (i) and (ii) is less than that
connecting (ii), (iii) and (iv).  This last observation can be attributed to the fact that in
case (i) the substrate is non-magnetic, whereas in case (ii) it is magnetic. In the absence
of magnetism, adsorption is stronger, and the points corresponding to cases (ii), (iii) and (iv)
would be shifted downwards, and one would have obtained a roughly straight line connecting
the points from the four cases considered by us; since these three points have, however, been shifted upwards by the presence of magnetism, the slope connecting (i) and (ii) is reduced. The main conclusion from this
graph is that the adsorption energy scales more-or-less linearly with effective
coordination; we note further that the Rh(111) surface has a larger $n_e$ (9.15) and
weaker $E_{ads}$ (-2.18 eV) than all four cases considered by us.

Similar conclusions
about co-adsorption can be drawn from the right-hand-side panel, where the
co-adsorption energy is found to vary monotonically with $n_e$. The variation
is less linear for this case, presumably because of changes in co-adsorption
geometry.

%%%%%%%%%%%%%%%%%%%%%%%%%%%%%%%%%%%%%%%%%%%%%%%%%%%%%%%%%%%%%%%%%%%%%%%
%%%%%%%%%%%%%%%%%%%%%%%%%%%%%%%%%%%%%%%%%%%%%%%%%%%%%%%%%%%%%%%%%%%%%%
The adsorption energy decreases as we go from (i) to (iv) due to two reasons:
(a) the strength of adsorption increases as the coordination is decreased, and
(b) the horizontal hollow geometry, which leads to stronger binding to the
substrate and a weaker and longer NO bond, becomes more favored at low
effective coordination. However, these effects would have been even more marked
if the substrate were to remain non-magnetic: in every case, we have seen that
NSP calculations (where magnetism is suppressed) point to stronger adsorption
than SP calculations.

For the VB geometry, the N-O bond length is found to be $\sim 1.2$ \AA, whereas
for the HH geometry, the bond length is increased to $\sim 1.3$ \AA, suggesting
that it should be easier to break the N-O bond in the latter case.

Our results show that strain and the presence of the oxide substrate contribute
to the increased strength of adsorption primarily through their effect on
$n_e$, and that both effects contribute to roughly the same extent. In the case
of the magnesia substrate considered by us, charge transfer plays a negligible
role, since the donation and back-donation mechanisms essentially cancel out.
This may not be true for more active oxide substrates, such as titania and
ceria.

\begin{figure} \epsfig{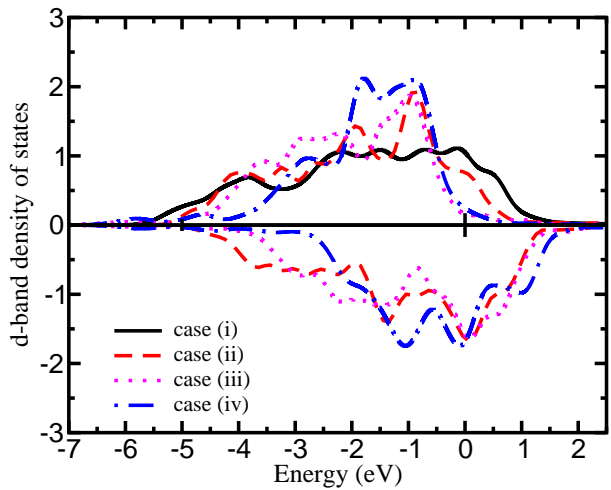}
\caption{(Color online) Rh $d$-band projected density of states (DOS)
for the four cases. Positive and negative values denote the DOS for spin-up
and spin-down electrons respectively.}
\label{dbandwidth} \end{figure}

It seems intuitively obvious that when Rh atoms have a lower effective
coordination, they will bind adsorbates more strongly, thus making it easier to
break bonds within the adsorbate. An alternative and equivalent way of looking
at this effect is to consider the effect of lower coordination on the density
of states. In Fig.~\ref{dbandwidth}, we show how the spin-polarized $d$-band
density of states of the surface Rh atoms changes as we go from case (i) to
case (iv) -- it can be seen that the lowering of $n_e$ results in a progressive
narrowing of the $d$-band. The $d$-bandwidth and position of the $d$-band
center have been shown to be a good predictor of catalytic activity.\cite{dband}

%%%%%%%%%%%%%%%%%%%%%%%%%%%%%%%%%%%%%%%%%%%%%%%%%%%%%%%%%%%%%%%%%%%%%%%
%%%%%%%%%%%%%%%%%%%%%%%%%%%%%%%%%%%%%%%%%%%%%%%%%%%%%%%%%%%%%%%%%%%%%%

\begin{figure} \epsfig{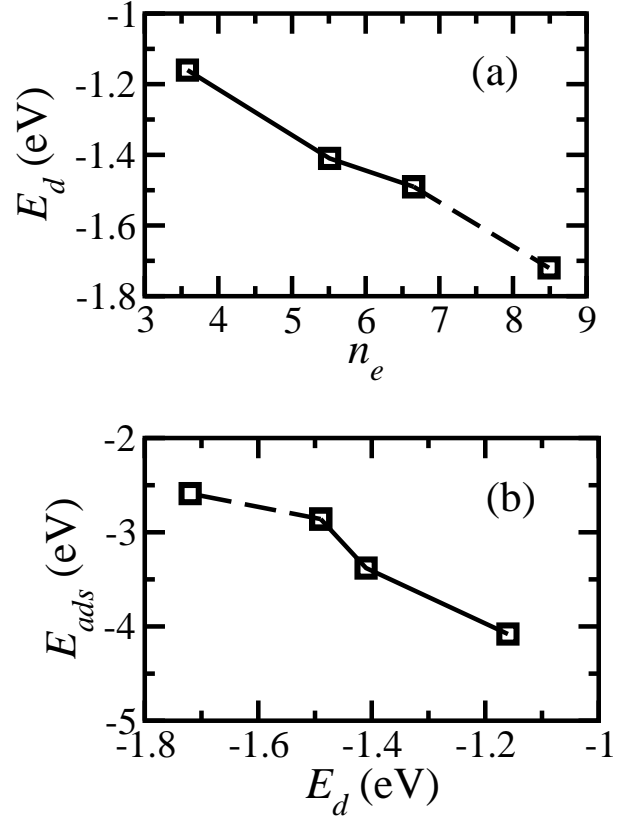}
\caption{Variation of (a) the position of the $d$-band center ($E_d$) with
effective coordination, and (b) the adsorption energy $E_{ads}$ with $E_d$. In
both cases, a monotonic and approximately linear relationship is observed. The
points plotted correspond to the lowest-energy configuration for each case.}
\label{dband_eads_ecoads_vs_neff} \end{figure}

In Fig.~\ref{dband_eads_ecoads_vs_neff}(a) we show how the position of the $d$-band center
shifts with effective coordination.  We see that as we go from case (i) to case
(iv), the reduction in $n_e$ is accompanied by an approximately linearly
proportional shift in the $d$-band center, bringing it closer to the Fermi
level. In Fig~\ref{dband_eads_ecoads_vs_neff}(b) we show that the adsorption energy varies in a
monotonic (and approximately linear) way with the $d$-band center.

%%%%%%%%%%%%%%%%%%%%%%%%%%%%%%%%%%%%%%%%%%%%%%%%%%%%%%%%%%%%%%%%%%%%%%%
%%%%%%%%%%%%%%%%%%%%%%%%%%%%%%%%%%%%%%%%%%%%%%%%%%%%%%%%%%%%%%%%%%%%%%

\section{Summary and Conclusions} We have studied the adsorption of NO and the
co-adsorption of N and O on strained and unstrained Rh surfaces with and
without the presence of an MgO substrate. Both strain and placing a monolayer
of Rh atoms on the oxide substrate lead to a significant lowering in the
effective coordination of surface Rh atoms; doing both [i.e., placing a
monolayer of Rh atoms pseudomorphically on an MgO(100) substrate] leads to the
largest decrease in effective coordination $n_e$. Further, both strain and the
presence of the substrate (either separately or together) have the effect of
making the surface Rh atoms magnetic. Every decrease in $n_e$ is accompanied by
(and, presumably, causes) a decrease in the $d$-bandwidth of Rh atoms, a shift
of the $d$-band center towards the Fermi level, and a strengthening of the
adsorption of NO and the co-adsorption of N and O.

We note that the effective coordination $n_e$ is a quantity that can be very
simply computed, especially if the structure is known -- it is not even
necessary to perform an {\it ab initio} density functional theory calculation
in order to compute it. Therefore it can serve as a simple guide or
rule-of-thumb in order to design systems where the strength of adsorption or
co-adsorption takes on a desired value.

Thus, lowering the effective coordination seems to be a good strategy to
increase the strength of adsorption and co-adsorption, and thus, conceivably,
lower the barrier to dissociation of the NO bond. One can think of several ways
of reducing effective coordination: e.g., by using rough surfaces, by placing
Rh atoms on an inert oxide substrate, and by using Rh nanocatalysts where the
Rh particles are sufficiently small so as to be significantly
under-coordinated. However, most of these strategies to reduce $n_e$ also favor
magnetism, which competes with bonding.\cite{NOclust}  This suggests that an optimal value of $n_e$ should be aimed
for, where the coordination is low enough so as to favor a significant
strengthening of Rh-NO bonds and weakening of the N-O bond, yet the adverse
effects of magnetism are not too evident.

We are in the process of computing dissociation barriers, in order to
verify our expectation that lowering effective coordination will also weaken
the N-O bond and thus catalyze the reduction of NO to N$_2$.

\bibliographystyle{apsrev} % This for APS and similar %<APS>
\bibliography{pghosh-manu}
\end{document}